\newif\ifproblem
\newif\iftimesok

\makeatletter
\def\IfStandaloneCheck{\def\next{aipcheck}
  \edef\currjob{\jobname}
  \edef\next{\meaning\next}
  \edef\currjob{\meaning\currjob}
  \ifx\currjob\next
    \expandafter\@firstoftwo
  \else
    \expandafter\@secondoftwo
  \fi
}
\makeatother

\ifx\documentclass\undefined
 \expandafter\stop
\else
\fi

\def\next#1/#2/#3\next{#1#2}
\ifnum\expandafter\next\fmtversion\next<199612 \relax
 \expandafter\stop
\else
 \ifnum\expandafter\next\fmtversion\next<199806 \relax
   \typein{* Type <return> to continue ...}
   \problemtrue
 \else
 \fi
\fi

\IfFileExists{aipproc.cls}
    {
     \typeout{* ... ok }
    }
    {
     \typeout{* ... not found! }
     \typeout{*}
     \typeout{* Sorry this is a fatal error:}
     \typeout{*}
     \typeout{* Before you can use the aipproc class you have to unpack}
     \typeout{* it from the documented source.}
     \typeout{*}
     \typeout{* Run LaTeX on the file 'aipproc.ins', e.g.,}
     \typeout{*}
     \typeout{* \space\space latex aipproc.ins}
     \typeout{*}
     \typeout{* or whatever is necessary on your installation to process}
     \typeout{* a file with LaTeX. This should unpack a number of files for you:}
     \typeout{*}
     \typeout{* aipproc.cls \space and \space aip-*.clo}
     \typeout{*}
     \typeout{* After that retry processing this guide.}
     \typeout{*}
     \stop
}

\IfFileExists{aipxfm.sty}
    {
     \typeout{* ... ok }
    }
    {
     \typeout{* ... not found! }
     \typeout{*}
     \typeout{* Sorry this is a fatal error:}
     \typeout{*}
     \typeout{* The aipxfm.sty file which is part of the aipproc distribution}
     \typeout{* must be installed in a directory which is searched by LaTeX.}
     \typeout{*}
     \typeout{* Please install this file and retry.}
     \typeout{*}
     \stop
}

\IfFileExists{aip-8s.clo}
    {
     \typeout{* ... ok }
    }
    {
     \typeout{* ... not found! }
     \typeout{*}
     \typeout{* Sorry this is a fatal error:}
     \typeout{*}
     \typeout{* The aip-8s.clo file which is part of the aipproc distribution}
     \typeout{* must be installed in a directory which is searched by LaTeX.}
     \typeout{*}
     \typeout{* Please install this file and retry.}
     \typeout{*}
     \stop
}

\IfFileExists{aip-8d.clo}
    {
     \typeout{* ... ok }
    }
    {
     \typeout{* ... not found! }
     \typeout{*}
     \typeout{* Sorry this is a fatal error:}
     \typeout{*}
     \typeout{* The aip-8d.clo file which is part of the aipproc distribution}
     \typeout{* must be installed in a directory which is searched by LaTeX.}
     \typeout{*}
     \typeout{* Please install this file and retry.}
     \typeout{*}
     \stop
}

\IfFileExists{aip-6s.clo}
    {
     \typeout{* ... ok }
    }
    {
     \typeout{* ... not found! }
     \typeout{*}
     \typeout{* Sorry this is a fatal error:}
     \typeout{*}
     \typeout{* The aip-6s.clo file which is part of the aipproc distribution}
     \typeout{* must be installed in a directory which is searched by LaTeX.}
     \typeout{*}
     \typeout{* Please install this file and retry.}
     \typeout{*}
     \stop
}

\IfFileExists{aip-arlo.clo}
    {
     \typeout{* ... ok }
    }
    {
     \typeout{* ... not found! }
     \typeout{*}
     \typeout{* Sorry this is a fatal error:}
     \typeout{*}
     \typeout{* The aip-arlo.clo file which is part of the aipproc distribution}
     \typeout{* must be installed in a directory which is searched by LaTeX.}
     \typeout{*}
     \typeout{* Please install this file and retry.}
     \typeout{*}
     \stop
}

\IfFileExists{fixltx2e.sty}
    {
     \typeout{* ... ok }
    }
    {
     \typeout{* ... not found, trying fix2col.sty instead ... }
     \typeout{*}
     \IfFileExists{fix2col.sty}
	 {
	  \typeout{* ... ok }
	 }
	 {
	  \typeout{* ... not found! }
	  \typeout{*}
	  \typeout{* Sorry this is a fatal error:}
	  \typeout{*}
	  \typeout{* Your LaTeX distribution contains neither fixltx2e.sty}
	  \typeout{* nor fix2col.sty.}
	  \typeout{*}
	  \typeout{* This means that it is either too old or incompletely}
	  \typeout{* installed.}
	  \typeout{*}
	  \typeout{* fixltx2e.sty is part of the standard LaTeX distribution}
	  \typeout{* since 1999; fix2col.sty is an earlier version of this}
	  \typeout{* package.}
	  \typeout{*}
	  \typeout{* Best solution is to get the latest LaTeX distribution.}
	  \typeout{* If this is impossible for you, download fix2col.sty.}
	  \typeout{* You can get this software from a CTAN host.}
          \typeout{* Refer to http://www.ctan.org and search for "fix2col".}
	  \typeout{*}
	  \typeout{* After you have updated your LaTeX distribution}
	  \typeout{* retry processing this guide.}
	  \stop
     }
}

\IfFileExists{fontenc.sty}
    {
     \typeout{* ... ok }
    }
    {
     \typeout{* ... not found! }
     \typeout{*}
     \typeout{* Sorry this is a fatal error:}
     \typeout{*}
     \typeout{* The fontenc package, which is part of standard LaTeX}
     \typeout{* (base distribution) has to be installed at the site to}
     \typeout{* run the aipproc class.}
     \typeout{*}
     \typeout{* The fact that it cannot be found either means that}
     \typeout{* this LaTeX release is too old or that it was installed}
     \typeout{* improperly.}
     \typeout{*}
     \typeout{* Please make sure that your version of LaTeX is okay}
     \typeout{* before attempting to use this class. The LaTeX distribution}
     \typeout{* contains the file "ltxcheck.tex" which can be used to}
     \typeout{* test the basic functionality and integrity of your installation.}
     \typeout{*}
     \stop
    }

\IfFileExists{calc.sty}
    {
     \typeout{* ... ok }
    }
    {
     \typeout{* ... not found! }
     \typeout{*}
     \typeout{* Sorry this is a fatal error:}
     \typeout{*}
     \typeout{* The calc package, which is part of standard LaTeX}
     \typeout{* (tool distribution) has to be installed at the site}
     \typeout{* to run the aipproc class.}
     \typeout{*}
     \typeout{* The fact that it cannot be found either means that}
     \typeout{* this LaTeX release is too old or that it was installed}
     \typeout{* only in parts.}
     \typeout{*}
     \typeout{* Please make sure that the tools distribution of LaTeX}
     \typeout{* is installed before attempting to use this class.}
     \typeout{*}
     \typeout{* (You might be able to get calc.sty separately for your}
     \typeout{* installation if you are unable to upgrade to a recent}
     \typeout{* distribution for some reason.)}
     \typeout{*}
     \stop
    }

\IfFileExists{varioref.sty}
    {
     \typeout{* ... ok }
     
    }
    {
     \typeout{* ... not found! }
     \typeout{*}
     \typeout{* Problem detected:}
     \typeout{*}
     \typeout{* The varioref package, which is part of standard LaTeX}
     \typeout{* (tool distribution) is not installed at this site.}
     \typeout{*}
     \typeout{* The fact that it cannot be found either means that}
     \typeout{* this LaTeX release is too old or that it was installed}
     \typeout{* only in parts.}
     \typeout{*}
     \typeout{* You can use the aipproc class without this package but }
     \typeout{* you cannot make use of the options "varioref" or "nonvarioref".}
     \typeout{*}
     \typeout{* Please also note that the aipguide.tex documentation}
     \typeout{* normally uses the "varioref" option to show its}
     \typeout{* effects (which  will now fail).}
     \typeout{*}
     \typein{* Type <return> to continue ...}
     \problemtrue

    }

\IfFileExists{times.sty}
    {
     \begingroup
       \RequirePackage{times}
       \global\expandafter\let\csname ver@times.sty\endcsname\relax    
       \long\def\next{ptm}
       \ifx\rmdefault\next
         \typeout{* ... ok }
         
         \endgroup
         \timesoktrue
       \else
         \endgroup
     \typeout{* ... obsolete! }
     \typeout{*}
     \typeout{* Serious problem detected:}
     \typeout{*}
     \typeout{* The times package, which is part of standard LaTeX}
     \typeout{* (psnfss distribution) is obsolete at this site.}
     \typeout{*}
     \typeout{* The fact that it contains incorrect code either means that}
     \typeout{* this LaTeX release is too old or that it was installed}
     \typeout{* only in parts with old files remaining!}
     \typeout{*}
     \typeout{* You can use the aipproc class without this package but}
     \typeout{* you have to specify the option "cmfonts" which result in}
     \typeout{* documents which are not conforming to the AIP layout specification!}
     \typeout{*}
     \typeout{* You can also try using the class in the following way:}
     \typeout{*}
     \typeout{* \space\space \string\documentclass[cmfonts]{aipproc}}
     \typeout{* \space\space \string\usepackage{times}}
     \typeout{* \space\space ...}
     \typeout{*}
     \typeout{* With luck this will result in Times Roman output but chances}
     \typeout{* are that you will get a larger number of error messages in}
     \typeout{* which case you have to remove the \string\usepackage declaration.}
     \typeout{*}
     \typein{* Type <return> to continue ...}
          \problemtrue
          
       \fi
    }
    {
     \typeout{* ... not found! }
     \typeout{*}
     \typeout{* Serious problem detected:}
     \typeout{*}
     \typeout{* The times package, which is part of standard LaTeX}
     \typeout{* (psnfss distribution) can not be found.}
     \typeout{*}
     \typeout{* The fact that this package cannot be found either means that}
     \typeout{* this LaTeX release is too old or that it was installed}
     \typeout{* only in parts!}
     \typeout{*}
     \typeout{* You can use the aipproc class without this package but }
     \typeout{* you have to specify the option "cmfonts" which result in}
     \typeout{* documents which are not conforming to the AIP layout specification!}
     \typeout{*}
     \typein{* Type <return> to continue ...}
     \problemtrue
     
    }

\iftimesok % don't bother testing other font options if times already
\IfFileExists{t1ptm.fd}
    {
     \typeout{* ... ok }
    }
    {
     \typeout{* ... not found, trying T1ptm.fd ... }
     \IfFileExists{T1ptm.fd}
          {
           \typeout{* ... ok }
          }
          {
           \typeout{* ... not found}
           \typeout{* Serious problem detected:}
           \typeout{*}
           \typeout{* The times package, which is part of standard LaTeX}
           \typeout{* (psnfss distribution) is available but the corresponding}
           \typeout{* .fd file (defining how to load Times Roman) is missing.}
           \typeout{*}
           \typeout{* The fact that this package is only partially installed}
           \typeout{* means that you LaTeX installation is unable to use Times}
           \typeout{* Roman fonts!}
           \typeout{*}
           \typeout{* You can use the aipproc class without this package but }
           \typeout{* you have to specify the option "cmfonts" which result in}
           \typeout{* documents which are not conforming to the AIP layout}
           \typeout{* specification!}
           \typeout{*}
           \typein{* Type <return> to continue ...}
           \problemtrue
           \timesokfalse
           
          }
    }

\fi

\newcommand\CheckFDFile[3]{%
  \typeout{*}
  \typeout{* Looking for #1#3.fd or #2#3.fd ... }
  \IfFileExists{#1#3.fd}
    {
     \typeout{* ... ok }
    }
    {
     \IfFileExists{#2#3.fd}
      {
       \typeout{* ... ok }
      }
      {\problemtrue
       \typeout{* ... not found! }
      }
    }
}

\iftimesok % don't bother testing other font options if Times already bad

\IfFileExists{mathptm.sty}
    {
     \typeout{* ... ok }
     \CheckFDFile{ot1}{OT1}{ptmcm}
     \CheckFDFile{oml}{OML}{ptmcm}
     \CheckFDFile{oms}{OMS}{pzccm}
     \CheckFDFile{omx}{OMX}{psycm}
     \ifproblem
      \typeout{*}
      \typeout{* Problem detected:}
      \typeout{*}
      \typeout{* The mathptm package, which is part of standard LaTeX}
      \typeout{* (psnfss distribution) was found but some or all of its}
      \typeout{* support files describing which fonts to load are missing!}
      \typeout{*}
      \typeout{*}
      \typeout{* The fact that this package is only partially installed}
      \typeout{* means that the mathptm package cannot be used!}
      \typeout{*}
      \typeout{* You can use the aipproc class without this package but }
      \typeout{* you have to specify the option "nomathfonts" so that}
      \typeout{* math formulas will be typeset using Computer Modern.}
      \typeout{*}
      \typein{* Type <return> to continue ...}
      \problemtrue
      
     \else
      \typeout{*}
      \typeout{* Looking for mathptmx.sty ... }
      \IfFileExists{mathptmx.sty}
       {
        \typeout{* ... ok }
        \CheckFDFile{ot1}{OT1}{ztmcm}
        \CheckFDFile{oml}{OML}{ztmcm}
        \CheckFDFile{oms}{OMS}{ztmcm}
        \CheckFDFile{omx}{OMX}{ztmcm}
        \ifproblem
	  \typeout{*}
	  \typeout{* Problem detected:}
	  \typeout{*}
	  \typeout{* The mathptmx package, which is part of standard LaTeX}
	  \typeout{* (psnfss distribution) was found but some or all of its}
	  \typeout{* support files describing which fonts to load are missing!}
	  \typeout{*}
	  \typeout{*}
	  \typeout{* The fact that this package is only partially installed}
	  \typeout{* means that the mathptmx package cannot be used!}
	  \typeout{*}
	  \typeout{* You can use the aipproc class without this package but }
	  \typeout{* you have to specify the option "mathptm" (no x) so that}
	  \typeout{* math formulas use the older version with upright greek letters.}
	  \typeout{*}
	  \typein{* Type <return> to continue ...}
	  \problemtrue
	  
        \fi
       }
       {
	\typeout{* ... not found! }
	\typeout{*}
	\typeout{* Problem detected:}
	\typeout{*}
	\typeout{* The mathptmx package, which is part of standard LaTeX}
	\typeout{* (psnfss distribution) can not be found.}
	\typeout{*}
	\typeout{* This is unfortunate but not a disaster as the older}
	\typeout{* version of the package "mathptm" (no x) seems to exist.}
	\typeout{*}
	\typeout{* You can use the aipproc class without this package but }
	\typeout{* you have to specify the option "mathptm" so that}
	\typeout{* math formulas use the older version with upright greek letters.}
	\typeout{*}
	\typein{* Type <return> to continue ...}
	\problemtrue
	
       }
      \fi
    }
    {
     \typeout{* ... not found! }
     \typeout{*}
     \typeout{* Problem detected:}
     \typeout{*}
     \typeout{* The mathptm package, which is part of standard LaTeX}
     \typeout{* (psnfss distribution) can not be found.}
     \typeout{*}
     \typeout{* The fact that this package cannot be found either means that}
     \typeout{* this LaTeX release is too old or that it was installed}
     \typeout{* only in parts!}
     \typeout{*}
     \typeout{* You can use the aipproc class without this package but }
     \typeout{* you have to specify the option "nomathfonts" so that}
     \typeout{* math formulas will be typeset using Computer Modern.}
     \typeout{*}
     \typein{* Type <return> to continue ...}
     \problemtrue
     
    }

\IfFileExists{mathtime.sty}
    {
     \typeout{* ... ok }
    }
    {
     \typeout{* ... not found! }
     \typeout{*}
     \typeout{* The mathime package can not be found.}
     \typeout{*}
     \typeout{* This is not a serious problem because this package is}
     \typeout{* only of interest if you own the commerical MathTime fonts.}
     \typeout{*}
     \typeout{* You can use the aipproc class without this package but }
     \typeout{* you cannot use the "mathtime" option of the class.}
     \typeout{*}
     \typein{* Type <return> to continue ...}
     \problemtrue
    }
\else
\fi % iftimesok

\IfFileExists{graphicx.sty}
    {
     \typeout{* ... ok }
    }
    {
     \typeout{* ... not found! }
     \typeout{*}
     \typeout{* Problem detected:}
     \typeout{*}
     \typeout{* The graphics package, which is part of standard LaTeX}
     \typeout{* (graphics distribution) can not be found.}
     \typeout{*}
     \typeout{* The fact that this package cannot be found either means that}
     \typeout{* this LaTeX release is too old or that it was installed}
     \typeout{* only in parts!}
     \typeout{*}
     \typeout{* You can use the aipproc class without this package but }
     \typeout{* you cannot use commands like \protect\includegraphics
                or \protect\resizebox}
     \typeout{* in this case.}
     \typeout{*}
     \typeout{* Please note that you will get a further error message below}
     \typeout{* about: "graphicx.sty not found" because the class will try}
     \typeout{* to load this package! Type return in response to that error.}
     \typeout{*}
     \typeout{* As a result the illustrations in aipguide will look strange.}
     \typeout{*}
     \typein{* Type <return> to continue ...}

     \gdef\resizebox##1##2{}
     \gdef\includegraphics{\textbf{graphics package missing:}}
     \problemtrue
    }

\IfFileExists{textcomp.sty}
    {
     \typeout{* ... ok }
    }
    {
     \typeout{* ... not found! }
     \typeout{*}
     \typeout{* Problem detected:}
     \typeout{*}
     \typeout{* The textcomp package, which is part of standard LaTeX}
     \typeout{* (base distribution) can not be found.}
     \typeout{*}
     \typeout{* The fact that this package cannot be found either means that}
     \typeout{* this LaTeX release is too old or that it was installed}
     \typeout{* only in parts!}
     \typeout{*}
     \typeout{* You can use the aipproc class without this package but }
     \typeout{* you will always get the error: "textcomp.sty not found"}
     \typeout{* because the class will try to load this package!}
     \typeout{* Type return in response to that error.}
     \typeout{*}
     \typein{* Type <return> to continue ...}

     \problemtrue
    }

\IfFileExists{url.sty}
    {
     \typeout{* ... ok }
    }
    {
     \typeout{* ... not found! }
     \typeout{*}
     \typeout{* Problem detected:}
     \typeout{*}
     \typeout{* The url package, which should be part of a good LaTeX}
     \typeout{* distribution, can not be found.}
     \typeout{*}
     \typeout{* Without this package you will not be able to use the \string\url}
     \typeout{* command. Try to download this package from a CTAN  host.}
     \typeout{* Refer to http://www.ctan.org and search for "url".}
     \typeout{*}
     \typein{* Type <return> to continue ...}

     \problemtrue
    }

\makeatletter

\IfFileExists{natbib.sty}
    {
     \IfStandaloneCheck
       {\begingroup
        \let\@listi\relax
        \let\thebibliography\@empty
        \let\bibstyle\@empty
        \RequirePackage{natbib}
        \@ifpackagelater{natbib}{1999/05/29}
	  {
           \typeout{* ... ok }
	  }{
           \typeout{* ... might be too old! }
           \typeout{*}
           \typeout{* Your version of the natbib package might be too}
           \typeout{* old to be usable. This class was designed to}
           \typeout{* work with the version 7.0 dated 1999/05/28}
           \typeout{*}
           \typeout{* If problems occur download a}
           \typeout{* recent version from a CTAN host.}
           \typeout{*}
           \typeout{* Refer to http://www.ctan.org and search for "natbib".}
           \typeout{*}
           \typein{* Type <return> to continue ...}

           \global\problemtrue
	  }
        \endgroup
        }{}
    }
    {
     \typeout{* ... not found! }
     \typeout{*}
     \typeout{* Serious problem detected:}
     \typeout{*}
     \typeout{* The natbib package, which should be part of a good LaTeX}
     \typeout{* distribution, can not be found.}
     \typeout{*}
     \typeout{* Without this package you will not be able to use certain}
     \typeout{* citation styles. See the aipguide documentation!}
     \typeout{*}
     \typeout{* Especially the layout for ARLO requires this package!}
     \typeout{*}
     \typeout{* Try to download this package from a CTAN  host.}
     \typeout{* Refer to http://www.ctan.org and search for "natbib".}
     \typeout{*}
     \typein{* Type <return> to continue ...}

     \problemtrue
    }

\makeatother

\ifproblem
\typein{* Type <return> to continue ...}
\else
\fi

\makeatletter
\IfStandaloneCheck
 {
\typeout{*}
\typeout{* This document only produces terminal output.}
\typeout{*}
\stop
 }
 {
\AtBeginDocument{\relax\ifx\xfm@address@loop\@undefined
  \typeout{***************************}
  \typeout{* Oooops ... you seem to have picked up an obsolete}
  \typeout{* aipproc.cls file from a previous installation!}
  \typeout{*}
  \typeout{* Please check that LaTeX finds the right one.}
  \typeout{*}
  \typeout{* Sorry have to give up ....}
  \typeout{***************************}
  \stop
 \fi}
 }
\makeatother

\documentclass[
    ,final            % use final for the camera ready runs
  ]
  {aipproc}

\layoutstyle{6x9}

\newcommand{\amu}{$a_\mu$}
\renewcommand{\to}{\ensuremath{\rightarrow}}
\renewcommand{\L}{\ensuremath{{\mathcal L}}}
\def\ifm#1{\relax\ifmmode#1\else$#1$\fi}
\def\sig{\ifm{\sigma}} \def\epm{\ifm{e^+e^-}}

\def\pic{\ifm{\pi^+\pi^-}} \def\gam{\ifm{\gamma}} 

\def\up#1{$^{#1}$}    
\def\ab{\ifm{\sim}}  \def\x{\ifm{\times}}    
  
\def\ie{{\it\kern-2pt i.\kern-.5pt e.\kern-2pt}}  
\def\dif{\hbox{d\kern.1mm}}  \def\deg{\ifm{^\circ}}
\def\up#1{$^{#1}$}  
\def\pt#1,#2,{\ifm{#1\x10^{#2}}}
 
\def\bye{\end{document}}

\begin{document}

\title{MEASUREMENT OF THE HADRONIC CROSS SECTION AT DA$\Phi$NE WITH
THE KLOE DETECTOR}

\author{The KLOE collaboration~\thanks{
%%%%%%%%%%%%%%%%%~\footnote{
%{\it M.~Adinolfi\rb,\hsa}
The KLOE Collaboration: A.~Aloisio,
F.~Ambrosino,
%{\it A.~Andryakov\mo,\hsa}
A.~Antonelli,
M.~Antonelli,
%{\it F.~Anulli\fr,\hsa}
C.~Bacci,
%G.~Barbiellini\t,\hsa 
%F.~Bellini\rc,\hsa
G.~Bencivenni,
S.~Bertolucci,
C.~Bini,
C.~Bloise,
V.~Bocci
% ...
%Y.~Wu\rlap,\kern.2ex\up{c,b}
%{\it Y.~G.~Xie\o,\hsa}
G.~Xu.
%P.~F.~Zema\pI,\hsa          
%{\it P.~P.~Zhao\o,\hsa          }
%Y.~Zhou\fr\hsa}
} 
\\presented by Achim G. Denig}{
  address={Universit\"at Karlsruhe, IEKP, Postfach 3640, 76021 Karlsruhe, Germany}
%, 
%  achim.denig@lnf.infn.it}
%           Institut f\"ur Experimentelle Kernphysik\\
%           Postfach 3640\\
%           D-76021 Karlsruhe\\
%           achim.denig@lnf.infn.it}
}
%
%\author{<author2>}{
%  address={<common address for author2 and author3>}
%}
%
%\author{<author3>}{
%  address={<common address for author2 and author3>}
%  ,altaddress={<author1 address>} % additional visiting address
%}

\begin{abstract}
We have measured the cross section $\sig(\epm\to\pic\gam)$ 
as a function of the $\pi^+\pi^-$ invariant mass, $M_{\pi\pi}$,
with the KLOE detector at DA$\Phi$NE ( $W=m_\phi=1.02$ GeV ).
%From the dependence of the cross section on $m(\pic)=\sqrt{W^2-2WE_\gamma}$, 
%where  $E_\gamma$ is the energy of the photon radiated from the initial 
%state, 
The photon in the above process is due to 
Initial State Radiation. Dividing by a theoretical radiator function, 
we obtain the cross section $\sig(\epm\to\pic)$ for the mass 
range $0.37<M^2_{\pi\pi}<0.93$ GeV$^2$. 
We extract the pion form factor and the hadronic contribution 
to the muon anomaly, $a_\mu$.
\end{abstract}

\maketitle

%%%%%%%%%%%%%%%%%%%%%%%%%%%%%%%%%%%%%%%%%%%%
%% MAINMATTER
%%%%%%%%%%%%%%%%%%%%%%%%%%%%%%%%%%%%%%%%%%%%

\section{Hadronic Cross Section at DA$\Phi$NE}

%The recent precision measurement of the muon anomaly \amu\ 
%at the Brookhaven National Laboratory~\cite{e821}
%has led to renewed interest in accurate measurements of the cross
%section for $e^+e^-$ annihilation into hadrons. 

%%%\subsection{Motivation}
{\bf Motivation}\\
Accurate measurements of the cross section for $e^+e^-$ annihilation 
into hadrons are of importance for an interpretation of the 
recent new precision measurement of the anomalous magnetic 
moment of the muon~\cite{e821}. 
Hadronic contributions 
to the photon spectral functions due to quark loops 
are not calculable in the framework of perturbative QCD. 
It is well known, however, that the 
%%%imaginary part of the 
hadronic piece of the spectral function 
is connected by unitarity to the cross section for \epm\to hadrons. 
A dispersion relation can thus be derived,
giving the contribution to \amu\ as an integral over the hadronic 
cross section, multiplied by an appropriate kernel. 
%An example of a first 
%complete estimate of the correction was given in 1985, 
%$\delta a_\mu^{\rm had} $=\pt707(6)(17),-10, \cite{kinoshita}.
The process \epm\to\pic\ below 1GeV is of special 
importance since it contributes to \ab60$\%$ to the total integral.
The most recent evaluation of the dispersion 
integral~\cite{davhoe}~\cite{davhoe2} gives
the following values for the muon anomaly, compared to the 
experimental value of the E821 collaboration:

\begin{tabular}[t]{ll} \hline
Theory using $\tau$ data    & $a_\mu^{theo} = (11\hspace{0.1cm}659\hspace{0.1cm}195.6 \pm 6.8)$x$10^{-10}$ \\
Theory using $e^+ e^-$ data & $a_\mu^{theo} = (11\hspace{0.1cm}659\hspace{0.1cm}180.9 \pm 8.0)$x$10^{-10}$ \\
Experiment BNL-E821         & $a_\mu^{exp} =  (11\hspace{0.1cm}659\hspace{0.1cm}203 \pm 8)$x$10^{-10}$ \\ \hline
\end{tabular}
\\
\\
The first value is obtained including hadronic $\tau$ decay data, assuming 
conservation of the vector current (CVC) and correcting for
isospin breaking effects. The second value uses 
$e^+e^-$ data only (see also~\cite{eidjeg}), in particular the
recent reanalysis~\cite{cmd2new} of the CMD-2 measurement~\cite{cmd2} of the $\pi^+\pi^-$ 
channel (0.6\% systematic error) in the energy range below 1 GeV.
The \epm\to\pic\ based result disagrees by \ab2 \sig\ 
with the BNL measurement, while the value using 
$\tau$ decay data is in agreement with the experimental value.
% with the experimental value for $a_\mu$.
Further independent hadronic cross section measurements are needed to clarify
the situation.
\\
\\
{\bf Initial State Radiation}\\
%%%\subsection{Initial State Radiation}
Particle factories such as DA$\Phi$NE or the B-factories typically 
operate at fixed centre-of-mass energies: $W=m_\phi$ in the case of DA$\Phi$NE.
Initial state radiation, ISR, is a complementary approach at particle factories
which allows studying \epm\to\ hadrons over the entire energy range 
(from 2$m_\pi$ to $W$). 
For a photon (energy $E_\gam$) radiated before annihilation of the 
\epm\ pair, the invariant mass of the  
\pic\ system is given by: $M^2_{\pi\pi}=s_\pi=W^2-2WE_\gam$. 
In general the \pic\gam\ and \pic\ cross section are related through:
\begin{equation}
s_\pi\,\frac{\dif\sigma(\pic\gam)}{\dif s_\pi}=
\sigma(\pic,\;s_\pi)\x H(s_\pi)
\label{eq:H}
\end{equation}
Eq. \ref{eq:H} defines the radiator function $H(s_\pi)$
which we obtain from the Monte Carlo program
PHOKHARA, a NLO generator for the $\pic\gam$ exclusive
final state~\cite{binner}~\cite{german}~\cite{czyz}~\cite{grzelinska}~\cite{grzelinska2}.
Our present analysis is based on the observation  of 
ref.\cite{binner}, that for small polar angles of the radiated photon, 
the ISR process dominates over the FSR process, which
is an indistinguishable background to our ISR approach. 
%We limit ourselves in the following to studying the 
In the following we present the cross section measurement of the
reaction \epm\to\pic\gam\ with $\theta_\gamma<15\deg$ or 
$\theta_\gamma>165\deg$~\footnote{For small $s_\pi$, the di-pion 
system is recoiling against the small angle photon, resulting in 
small angle pion tracks which cannot be detected in the KLOE drift
chamber.
We are therefore limited to measuring \sig(\pic) for $s_\pi>$0.3 GeV$^2$.
A complementary analysis in which photons are selected at large
angles is in progress.
In this case the photon can be tagged in the electromagnetic 
calorimeter and the kinematical acceptance allows us to measure 
events down to the $2\pi$ threshold.}. 
No explicit photon detection is done since the KLOE electromagnetic
calorimeter (EmC) does not cover angles smaller than 20$^o$. 
We cut on the di-pion production angle, $\theta_{\pi\pi}$, 
which is calculated from the momenta of the two charged tracks. 
If only one photon is emitted, the following relation holds exactly: 
$\theta_{\gamma}=180^\circ-\theta_{\pi\pi}$.
As we will show in the following, 
an efficient and almost background free signal selection can be 
obtained without photon tagging.

\section{ Analysis of $\pi^+\pi^-\gamma$ events}

We have analyzed KLOE data taken in 2001 with an integrated 
luminosity of $140$pb$^{-1}$. 
After fiducial volume and selection cuts we collect \ab\pt1.5,6, events. 
To obtain the cross section, we subtract the residual 
background from this spectrum and divide by the selection
efficiency and the integrated luminosity.
%\footnote{
%We have not unfolded data for mass resolution effects.
%MC studies have shown that the effect of detector smearing is small.
%} 
%:
%\begin{equation}
%{\Delta\sig_{\pic\gam}\over\Delta s_\pi}=%
%{\Delta N_{\pic\gam}\over\L\;\Delta s_\pi}=%
%{\Delta N_{\rm obs}-\Delta N_{\rm bckgnd}\over\L\;\Delta s_\pi}\x%
%{1\over\epsilon_{\rm cuts}}
%\label{eq:sighad}
%\end{equation}
\\
We briefly comment in the following on the individual analysis items.
Further details can be found in~\cite{aachen}. 
For a detailed description of the KLOE detector, which consists of a 
high resolution tracking detector ($\sigma_{p_T} / p_T \leq 0.4\%$) 
and an electromagnetic calorimeter ($\sigma_E / E = 5.7 \% /  \sqrt{E({\rm GeV})}$),
we refer to ref.~\cite{dc}~\cite{emc}.

\begin{itemize} 
\item {\it Detection of two charged tracks}: with polar angle larger 
than $50^\circ$, coming from a vertex in the fiducial volume $R<8$ cm, 
$|z|< 7$ cm. The cuts on the transverse momentum $p_{T} > 160$ MeV or on the 
longitudinal momentum $|p_{z}| > 90$ MeV reject tracks spiralizing along the 
beam line, ensuring good reconstruction conditions. 
The probability to reconstruct a vertex in the drift chamber is $\sim 98\%$ and 
has been studied with $\pi^+ \pi^- \pi^0$ and $\pi^+ \pi^-$ data.
%, selected using the calorimeter only. 
%This efficiency and the track reconstruction efficiency are going to be 
%also evaluated using $\pi^{+}\pi^{-}\pi^{0}$ events selected by detecting 
%$\pi^{0}$ in the electromagnetic calorimeter. 
%
\item {\it Identification of pion tracks}: A Likelihood Method 
(calibrated on real data), using the time of flight of the particle 
and the shape of the energy deposit in the electromagnetic calorimeter, 
has been developed to reject $e^{+}e^{-} \rightarrow e^+e^-\gam$ background. 
Background from 
$e^+e^-\gam$ events is drastically reduced like that. A control sample of 
$\pi^{+}\pi^{-}\pi^{0}$ has been used to study the behaviour of 
pions in the electromagnetic calorimeter and to evaluate the selection 
efficiency ($>98\%$) for signal events.

\item {\it Background subtraction}: $\mu^+\mu^-\gam$ and $\pi^{+}\pi^{-}\pi^{0}$ 
events are rejected by a cut in a kinematic variable called track mass, $m_{\rm trk}$  
This variable is calculated from the reconstructed pion momenta, 
$\vec{p}_{+}$, $\vec{p}_{-}$, applying 4-momentum conservation under the 
hypothesis that the final state consists of two particles 
with the same mass and one photon. 
For such $\epm\to x^+x^-\gamma$ events, the  value of $m_{\rm trk}$ peaks at 
$m_\pi,\ m_\mu,\ m_e$ for $x=\pi,\mu,e$ respectively, thus allowing a 
selection of signal events.
The density distribution of the two track events in the 
[$s_\pi,\ m_{\rm trk}$] 
plane is very effective for separating signal from background.
The final event selection is defined by: 
$(m_{\rm trk}>120) \cap (m_{\rm trk}<250-105\sqrt{1-(s_\pi/850000)^2}) 
\cap (m_{\rm trk}<220)$, 
all units in MeV. 
%The residual amount of background in the signal region is 
%obtained by fitting the measured \mtrk\ spectrum with the Monte Carlo  
%spectra for \pic\,\eeg\ and \mmg\ events. 
%Residual background from \ppp\ events appears at higher \mtrk\ values and is 
%obtained by fitting the missing mass distribution which is peaked at
%$m^2_{miss} = m(\po)$ for \ppp\ events.

\item {\it Luminosity Measurement}:
The  integrated luminosity is measured with the KLOE detector 
using large angle Bhabha (LAB) events. The effective Bhabha cross section 
at large angles 
($55^o < \theta_{+,-} < 125^o$) is about $430$ nb.  
%This cross section is large enough so that the statistical error
%on the luminosity measurement is negligible.
%The number of LAB candidates, $N_{LAB}$, is counted and normalized to the 
%effective Bhabha cross section.
%, obtained by Monte Carlo: 
%\begin{equation}
%\int{\mathcal L}dt =
%\frac{N_{LAB}(\theta_i)}{\sigma_{LAB}^{MC}(\theta_i)}
%\cdot (1-\delta_{Bkg})
%\end{equation}
%The precision of this measurement depends on the correct inclusion of higher order 
%terms in computing the Bhabha cross section and on an accurate simulation of the
%process in the detector simulation program. 
For the computation of the radiative corrections  
we use two independent Bhabha event generators: BHAGENF~\cite{berends}~\cite{drago}) 
and BABAYAGA (\cite{babayaga}). For each generator a systematic error of $0.5\%$ is 
quoted by the authors. The two generators agree to better than 0.2\% with each other.
%LAB events are selected with cuts on variables which are well simulated by the KLOE 
%Monte Carlo simulation.
%The electron and positron polar angle cuts, 55\deg$\,<\theta_{+,-}<\,$125\deg, is based 
%on the calorimeter clusters, while the energy cuts, $E_{+,-}>400$ MeV, is based on 
%drift chamber information. 
%The background from $\mu^+ \mu^- (\gamma)$, $\pi^+ \pi^- (\gamma)$ and 
%$\pi^+ \pi^- \pi^0$ 
%events is well below $1\%$ and is subtracted. 
All selection efficiencies (trigger, EmC 
cluster, DC tracking) are $>99\%$ and are well 
reproduced by the detector simulation program. 
We also obtain excellent agreement between the experimental distributions 
($\theta_{+,-}$, $E_{+,-}$) and those obtained from Monte Carlo simulation.
%The background from $\mu^+ \mu^- (\gamma)$, $\pi^+ \pi^- (\gamma)$ 
%and $\pi^+ \pi^- \pi^0$ 
%events is well below $1\%$ and is subtracted.
The experimental uncertainty in the acceptance due to all these effects is 0.4\%. 
We assign a total systematic error for the luminosity of 
$\delta\L=0.5\%_{\rm th}\oplus0.4\%_{\rm exp}$. 
%An independent check of the luminosity measurement is obtained using \epm\to\gam\gam\ 
%events. We find  agreement to within $0.2\%$.
%
%\item {\it Unfolding the mass resolution}: 
%In order to obtain $\dif\sigma_{\pi \pi \gamma}/\dif s_\pi$ as a function of the 
%true $s_\pi$ value, we unfold the mass resolution from the measured $s_\pi$ distribution.
%The measured value of $s_{\pi,{\rm obs}}$ is related to the true value by a resolution 
%matrix {\bf G}($s_{\pi,{\rm true}}-s_{\pi,{\rm obs}}|s_{\pi,{\rm true}}$). This matrix 
%has been generated by Monte Carlo simulation. 
%The resolution matrix (bin width of 0.01GeV$^2$) is 
%nearly diagonal with $75\%$ of the events in the diagonal, 
%$11\%$ in each of the first once-off-diagonal elements and only $3\%$ 
%in the remaining elements.
%Unfolding is obtained by applying the inverse matrix {\bf G}$^{-1}$ to the data.
\end{itemize}

The contribution of the several analysis items to the
total systematic error is shown in table~\ref{tab:syseff1}.
The value for the total systematic error of 1.2\% is preliminary and expected
to be reduced to \ab 1.0\% soon.
We have not unfolded data for mass resolution effects.
MC studies have shown that the effect of detector smearing is small.

\begin{table}
%\begin{center}
\begin{tabular}{||l|c||}
\hline
Acceptance & 0.3\% \\%[-2mm]
Trigger + Offline Reconstruction Filter & 0.6\% \\%[-2mm]
Tracking & 0.3\% \\%[-2mm]
Vertex & 0.7 \% \\%[-2mm]
Particle ID Estimator (Likelihood) & 0.1 \% \\%[-2mm]
Track Mass & 0.2 \% \\%[-2mm]
Background subtraction & 0.5 \% \\%[-2mm]
%Unfolding & 0.6 \% \\
\hline
Total experimental systematics & 1.2 \% \\
\hline
%\hline
%luminosity & 0.6 \% \\[-2mm]
%Vacuum Polarization &  0.2 \% \\
%\hline
%Total theor. systematics & 0.7 \% \\
%\hline
%\hline
%FSR resummation & 2.0 \% \\
%\hline
%\hline
\end{tabular}
\caption{List of systematic uncertainties.}
\label{tab:syseff1}
%\end{center}
\end{table}

\section{Results}

\begin{figure}[t]
  \includegraphics[height=.35\textheight]{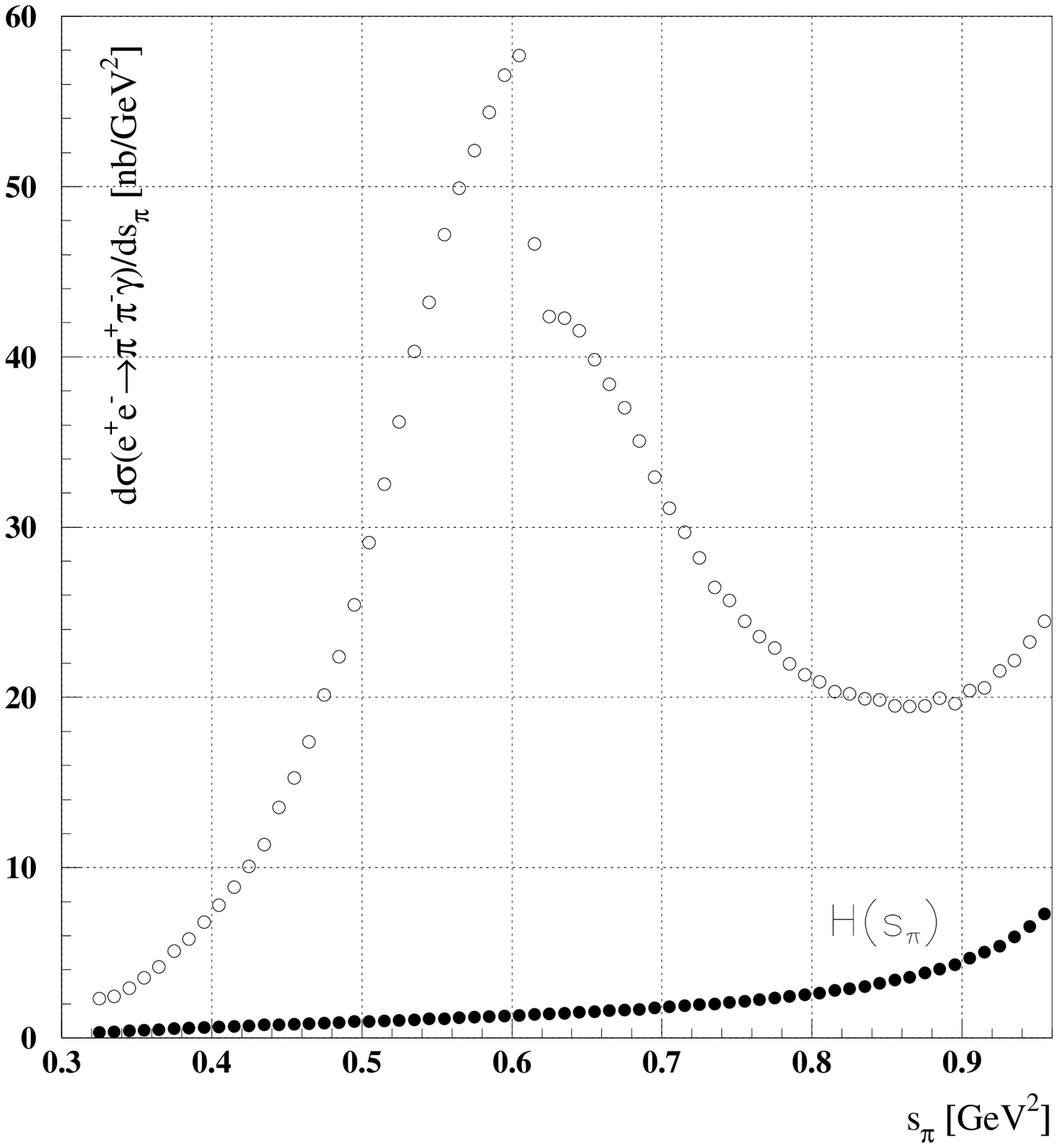}
  \includegraphics[height=.35\textheight]{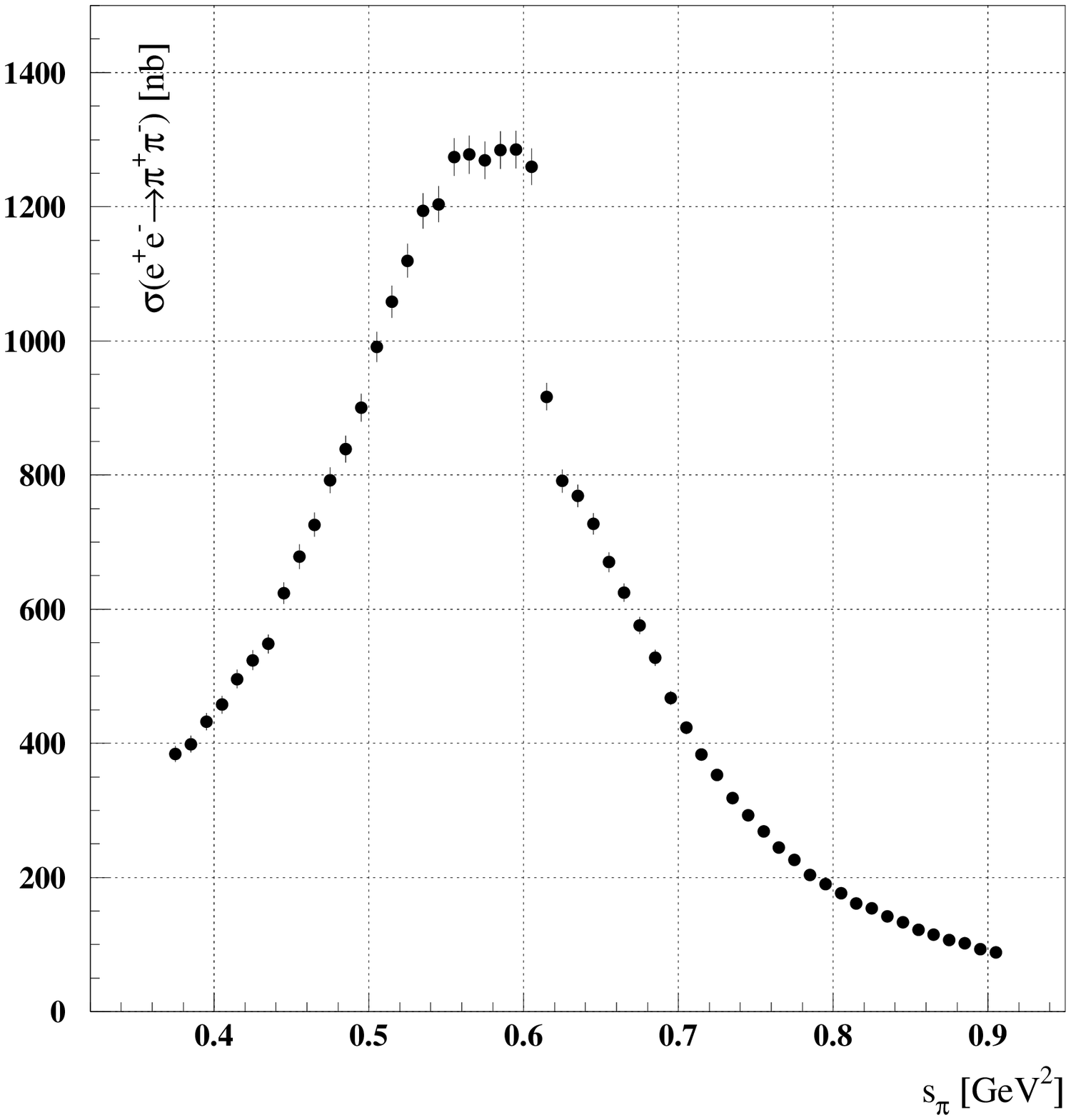}
  \caption{Left: Cross section for $e^+e^-\rightarrow \pi^+\pi^-
  \gamma$. The radiator $H(s_\pi)$, is also shown.
  Right: Bare cross section for $e^+e^-\rightarrow \pi^+\pi^-$.}
  \label{fig:sigppg}
\end{figure}

The result of our cross section measurement 
for $e^+ e^- \rightarrow \pi^+ \pi^- \gamma$ 
is shown in fig.~\ref{fig:sigppg}.
According to eq.~\ref{eq:H}
the radiator function $H(s_\pi)$ is needed 
in order to extract $\sigma(e^+ e^- \rightarrow \pi^+ \pi^-)$.
We obtain $H(s_\pi)$ from PHOKHARA, setting $F_\pi(s_\pi)=1$ and 
switching off the vacuum polarization of the intermediate photon
in the generator. 
The radiator $H(s_\pi)$ is also shown in fig.~\ref{fig:sigppg}, left.
The H-function is theoretically known with a precision of \ab 0.5\%. 
We take this value (together with the error on luminosity) 
as the theory error.
\\
%\subsection{FSR and Vacuum Polarization Corrections}
\\
{\bf FSR and Vacuum Polarization Corrections}\\
%If the hadronic cross section is used to evaluate the hadronic 
%contribution to the muon anomaly, 
Two further radiative corrections have to be applied to our data 
before evaluating the hadronic contribution to the muon anomaly
(see ref.~\cite{davhoe} for details).
The cross section has to be
the {\it bare} cross section, i.e. vacuum polarization corrections
must be subtracted. This can be done 
by correcting the cross section for the running of $\alpha$.
%, as follows:
%\begin{equation}
%\sigma_{\rm bare} = \sigma_{\rm dressed} \left( \frac{\alpha(0)}{\alpha(s)}
%\right)^2
%\end{equation}
%
%
\\
The second correction deals with the treatment of FSR events. In our 
$\pic\gam$ cross section measurement a part of multiphoton events 
(more than 1 hard photon) are removed by the $m_{trk}$ cut. 
%The efficiency of this cut is evaluated by Monte Carlo using a
%version of PHOKHARA, in which photons from ISR {\it and} FSR are 
%not simulated simultaneously. 
The efficiency of this cut is evaluated by Monte Carlo using a 
recently published new version of PHOKHARA~\cite{grzelinska2}, in which
simultaneously photons from ISR and FSR are simulated.
A preliminary study shows, that 
the cross section is changed by less than 2\% due to this
next-to-leading-order FSR effect.  
%The final corrections are under study and final results
%are expected soon. 
We are working out the final corrections to be applied to data.
For now, we apply an error of 1\% as a conservative 
estimate of the uncertainty due to FSR corrections.
\\
%\subsection{Hadronic contribution to the muon anomaly}
\\
{\bf Hadronic contribution to the muon anomaly}\\
We have evaluated the hadronic two-pion contribution to the muon anomaly, 
$a_\mu^{\pi\pi}$, by inserting our measured {\it bare} cross section
$e^+ e^- \rightarrow \pi^+\pi^-$ into the dispersion integral.
In order to compare our result with the CMD-2 result we 
cover the same energy interval as CMD-2. The 
preliminary KLOE result (in 10$^{-10}$ units) is 
in good agreement with the CMD-2 value as can be seen in the following table
\footnote{These numbers are the outcome of an updated analysis, presented at
SIGHAD03 (October 8-10,2003) and supersede the result presented at 
HADRON2003: 
$a_\mu^{\pi\pi}=374.1\pm1.1_{\rm stat}\pm 5.2_{\rm syst}\pm 2
.6_{\rm theo}\left.\; ^{+7.5}_{-0.0}\right|_{\rm FSR}$}
:

\begin{tabular}[t] {cl} \hline
KLOE preliminary & $a_\mu^{\pi\pi}=378.4\pm0.8_{\rm stat}\pm 4.5_{\rm syst}\pm 3.0_{\rm theo}\pm3.8_{\rm FSR}$ \\
CMD-2 reanalysis 2003~\cite{cmd2new}& $a_\mu^{\pi\pi}=378.6\pm2.7_{\rm stat}\pm 2.3_{\rm syst+theo}$\\ \hline
\end{tabular}
\\
The actual systematic error 
of $1.2\%$ will be reduced down to $\simeq 1\%$ in the very near future.
Also the error on FSR will be further reduced. 
%As discussed above, no FSR corrections have yet been applied.
The statistical error is negligible.
%This error  will be substantially reduced once the new PHOKHARA 
%version will be inserted into the KLOE MC.
%The central value of our result in the region $0.37<s_\pi<0.93$ is 
%slightly larger than the one obtained by CMD-2. 
%The discrepancy is, at the moment, not significant (0.5 $\sigma$).
%This difference is, however, not equally distributed in $s_\pi$, 
%as summarized below.
\\
Of special interest is the energy region above the $\rho$ peak 
because large discrepancies between evaluations based on 
$\tau$ data and \epm\to\pic\ data are seen here.
For this reason we have calculated $a_\mu^{\pi\pi}$ in the range
below (<0.65 GeV$^2$) and above (>0.65 GeV$^2$) the $\rho$ peak 
and compare the values with the ones from CMD-2\footnote{This 
is our evaluation of $a_\mu^{\pi\pi}$~[CMD-2], based on the values 
tabulated in ref.~\cite{cmd2new}}:
$$\vbox{\halign{\hfil#\hfil\qquad&\hfil # \hfil\qquad&\hfil#\hfil\cr
$s_\pi$ / GeV\up2&$a_\mu^{\pi\pi}$ [KLOE]&$a_\mu^{\pi\pi}$ [CMD-2]\cr
0.37 -- 0.65&$309.4\pm6.0$ & $308.5\pm2.8$\cr
0.65 -- 0.93&$68.8\pm1.2 $ & $72.2\pm0.7$ \cr}}$$

Our data are in good agreement with CMD-2 below the $\rho$ peak.
For the energy range $0.65<s_\pi<0.93{\rm GeV}^2$ our data 
are lower than CMD-2. Final corrections for FSR are still missing.
We conclude that our data confirm the 
results from \epm\to\pic\ and the discrepancy with respect to $\tau$ data.
%
%Already before these corrections we are able to conclude 
%that our data confirm the discrepancy ( \ab10\% ) seen at high energies 
%between $\tau$ data and \epm\to\pic\ results.

\end{document}

%%%\endinput